**On the Role of Tsallis Entropy Index for Velocity Modelling in Open Channels**

Manotosh Kumbhakar[1], Rajendra K. Ray[1], Koeli Ghoshal[2] and Vijay P. Singh[3, 4, 5]

[1]School of Basic Sciences, Indian Institute of Technology Mandi, Mandi 175005, India.

[2]Department of Mathematics, Indian Institute of Technology Kharagpur, Kharagpur 721302, India.

[3]Department of Biological & Agricultural Engineering, Texas A & M University, College Station, TX, 77843-2117, USA.

[4]Zachry Department of Civil & Environmental Engineering, Texas A & M University, College Station, TX, 77843-2117, USA.

[5]National Water Center, UAE University, Al Ain, UAE

[*]**Corresponding author details:**

Name – Rajendra K. Ray

Email – rajendra@iitmandi.ac.in

Contact No. - +91-1905-267041

**Abstract**

Following the work on Shannon entropy together with the principle of maximum entropy, Luo & Singh (J. Hydrol. Eng., 2011, 16(4): 303-315) and Singh & Luo (J. Hydrol. Eng., 2011, 16(9): 725-735) explored the concept of non-extensive Tsallis entropy for modelling velocity in open channels. Later, the idea was extended by Cui & Singh (J. Hydrol. Eng., 2013, 18(3): 331-339; 2014, 19(2): 290-298) by hypothesizing an accurate cumulative distribution function (CDF). However, these studies estimated the entropy index through a data-fitting procedure and the values of the index were different for different studies. The present study investigates the role of Tsallis entropy index for modelling velocity in open channels using the method of moments, based on conservation of mass and momentum. It is found that the entropy index depends on the normalized mean velocity and the momentum coefficient. In addition to the physical meaning of the index, it is also found that the modified velocity profile significantly improves for both wide and narrow channels, as shown by small predicted velocity errors. The proposed approach may be further employed for other open channel flow problems, such as sediment concentration, and shear stress distribution.

**Keywords:** Tsallis entropy, Entropy index, Open channel flow, Method of moments, Velocity Distribution.

1. Introduction

The velocity distribution in open channels is of fundamental importance in hydraulic modelling. Both deterministic and statistical models of velocity distribution have been developed in the literature. Deterministic models are mostly empirical, based on experimental data. Some of the statistical models employ the entropy theory which has long been applied in hydraulics (*Singh 2014, 2016*). Using Shannon entropy and the principle of maximum entropy (POME), *Chiu (1987)* derived velocity distributions in open channels. Following his work, different researchers applied POME to Shannon entropy for studying hydraulic problems (*Singh 2014, Kundu 2017, Kumbhakar* *et al.* 2017, 2018). There is another entropy, called Tsallis entropy, which is non-extensive in nature and is a generalization of Shannon entropy containing an additional parameter. Tsallis entropy takes on maximum value in case of equiprobability and is pseudo-additive for independent subsystems. Because of these and other essential properties, it has been receiving a great deal of attention, especially in physics (details can be found here: http://tsallis.cat.cbpf.br/biblio.htm).

Tsallis entropy was first employed by *Koutsoyiannis (2005a, b)* who characterized the stochastic behavior of hydrological processes. *Keylock (2005)* derived the flood recurrence intervals using a $q$-exponential distribution based on Tsallis entropy. In hydraulics, *Singh and Luo (2011)* and *Luo and Singh (2011)* explored the concept of non-extensive entropy for studying the velocity distributions in wide and narrow open channels, respectively. Their works showed some advantages over the Shannon entropy-based velocity profile given by *Chiu (1987)* in predicting the velocity distribution, especially near the channel bed. Later, the idea was further extended by *Cui and Singh (2013, 2014)* by hypothesizing an accurate CDF in the space domain. Besides studies on velocity, Tsallis entropy theory was successfully applied to other problems of hydraulics and hydrology also (*Singh 2016*). However, the mentioned works estimated the value of the entropy index using a data-fitting procedure without any physical justification. In general, except for a few cases such as *Lyra and Tsallis (1998)* and *Conroy and Miller (2015)*, the Tsallis entropy index seems to be an adjustable parameter that can lead to a proper distribution through validation with experimental data (*Tsallis 1999*).

Therefore, the objective of the present study is to investigate the role of Tsallis entropy index for modelling velocity in open channels and determine the improvement in the derived velocity profile over the existing one using selected sets of experimental and field data.

## 2. Methodology

### 2.1. *Probability Distribution of Velocity*

An open channel flow of depth $D$ and width $B$ is considered where the time-averaged streamwise velocity $u$ is distributed in the vertical ($y$) and transverse ($z$) directions, respectively. The normalized velocity $\hat{u} = u/u_{max}$, $u_{max}$ being the maximum velocity in a given cross-section, is assumed to be a random variable. Then, the Tsallis entropy (*Tsallis* 1988) of $\hat{u}$, having a PDF $f(\hat{u})$, can be written in continuous (differential) form as

$$H_q[f(\hat{u})] = \frac{1}{q-1} \int_{\hat{u} \in \Theta} f(\hat{u}) \left[1 - (f(\hat{u}))^{q-1}\right] d\hat{u} \tag{1}$$

where $\Theta = [0,1]$ is the domain of $\hat{u}$, $q$ is the real parameter often called entropy index. It is straightforward to check that as $q \to 1$, Eq. (1) recovers the Shannon entropy. For all $q$, the Tsallis entropy attains its maximum in the case of uniform distribution for $f(\hat{u})$. Moreover, the function $H_q$ is concave for $q > 0$ and convex for $q < 0$ (*Plastino and Plastino 1999*).

The objective is to determine the Tsallis distribution by applying the principle of maximum entropy to Tsallis entropy Eq. (1), subject to the specified constraints. If observations on velocity are available, the constraints can be prescribed as follows. The PDF $f(\hat{u})$ must obey the total probability rule, i.e.,

$$\int_{\hat{u} \in \Theta} f(\hat{u}) d\hat{u} = 1 \tag{2}$$

Eq. (2) is generally known as the normalization constraint. The second constraint is formulated by noting that the area integral of $u$ in a given cross-section is equal to discharge or flow rate, i.e.,

$$Q = \int_A u dA = u_{max} \int_A \hat{u} dA = u_{max} A \bar{\hat{u}} = u_{max} A \int_{\hat{u} \in \Theta} \hat{u} f(\hat{u}) d\hat{u}$$

$$\Rightarrow \int_{\hat{u} \in \Theta} \hat{u} f(\hat{u}) d\hat{u} = \bar{\hat{u}} = \frac{Q}{A u_{max}} \tag{3}$$

where $Q$ is the discharge, $A$ is the cross-sectional area, and $\bar{\hat{u}} = \frac{\bar{u}}{u_{max}}$ is the normalized mean velocity. It may be noted that Eq. (3) is the first statistical moment of the random variable $\hat{u}$ and represents the hydrodynamic transport of mass by the flow through the cross-section of a channel. Similarly, other constraints may be imposed; however, we restrict our study to mass conservation constraint only, because previous studies have shown that this constraint suffices for deriving velocity profiles.

The Tsallis entropy given by Eq. (1) can now be maximized subject to constraints Eqs. (2) and (3). To that end, introducing the Euler-Lagrange method of calculus of variations, the Lagrangian function can be constructed as

$$L(f,\lambda) = \frac{1}{q-1}\int_{\hat{u}\in\Theta} f(\hat{u})\left[1-(f(\hat{u}))^{q-1}\right]d\hat{u} + \lambda_0\left(\int_{\hat{u}\in\Theta} f(\hat{u})d\hat{u} - 1\right) + \lambda_1\left(\int_{\hat{u}\in\Theta}\hat{u}f(\hat{u})d\hat{u} - \bar{\hat{u}}\right) \quad (4)$$

Using Euler-Lagrange equation $\frac{\partial L}{\partial f} - \frac{\partial}{\partial \hat{u}}\left(\frac{\partial L}{\partial f'}\right) = 0$ for Eq. (4), the PDF can be obtained as

$$f(\hat{u}) = \left(\frac{q-1}{q}\left[\frac{1}{q-1} + \lambda_0 + \lambda_1\hat{u}\right]\right)^{\frac{1}{q-1}} \quad (5)$$

Substitution of PDF Eq. (5) into the constraints yields the following system of equations:

$$\frac{1}{\lambda_1}\left(\frac{q-1}{q}\right)^{\frac{q}{q-1}}\left[\left(\lambda_0 + \lambda_1 + \frac{1}{q-1}\right)^{\frac{q}{q-1}} - \left(\lambda_0 + \frac{1}{q-1}\right)^{\frac{q}{q-1}}\right] = 1 \quad (6)$$

$$\frac{1}{\lambda_1^2}\left(\frac{q-1}{q}\right)^{\frac{1}{q-1}}\left[\frac{q-1}{2q-1}\left\{\left(\lambda_0 + \lambda_1 + \frac{1}{q-1}\right)^{\frac{2q-1}{q-1}} - \left(\lambda_0 + \frac{1}{q-1}\right)^{\frac{2q-1}{q-1}}\right\}\right.$$
$$\left. - \frac{q-1}{q}\left(\lambda_0 + \frac{1}{q-1}\right)\left\{\left(\lambda_0 + \lambda_1 + \frac{1}{q-1}\right)^{\frac{q}{q-1}} - \left(\lambda_0 + \frac{1}{q-1}\right)^{\frac{q}{q-1}}\right\}\right]$$
$$= \bar{\hat{u}} \quad (7)$$

Also, the cumulative distribution function (CDF) can be obtained from Eq. (5) as

$$F(\hat{u}) = Prob(\hat{U} \leq \hat{u}) = \int_0^{\hat{u}} f(\hat{u})d\hat{u}$$

$$= \frac{1}{\lambda_1}\left(\frac{q-1}{q}\right)^{\frac{q}{q-1}}\left\{\left(\lambda_0 + \lambda_1\hat{u} + \frac{1}{q-1}\right)^{\frac{q}{q-1}} - \left(\lambda_0 + \frac{1}{q-1}\right)^{\frac{q}{q-1}}\right\} \quad (8)$$

Eq. (8) is the Tsallis entropy-based CDF which is obtained by the maximum entropy principle. One of the objectives is to derive the streamwise velocity equation along vertical and transverse directions. To that end, the connection between the velocity and the space domain is made in what follows.

*2.2. Connection with Space Domain*

Based on the aspect ratio, i.e., the ratio of channel width $B$ to flow depth $D$, an open channel is classified as wide and narrow channels. If $B/D < 5$ then the channel is narrow, for $B/D > 10$, it is wide, and for other cases, it depends on the condition of the surface roughness. In addition to the primary flow, strong secondary currents exist in narrow channels due to which the maximum velocity occurs below the free surface, which is commonly known as *dip phenomenon (Chow 1959)*. Therefore, in order to have a unique, one-to-one relationship between the space coordinate and a value of velocity, *Chiu (1989)* constructed the following generalized coordinate

$$\Psi = Y(1-Z)^m \exp(mZ - Y + 1) \qquad (9)$$

in which $Y = \frac{y}{D+h}$ and $Z = \frac{|z|}{B_i}$, where $D$ is the maximum flow depth, $h$ is the vertical position of maximum velocity below the water surface (in downward direction), $B_i$ for $i = 1$ or $2$ represents the transverse distance between the y-axis and either the left or right bank of a channel cross-section, and $m$ is a parameter which characterizes the shape of velocity profiles.

The velocity $\hat{u}$ now becomes monotonically increasing with respect to $\Psi$-cordinate over the entire domain $[\Psi_{min}, \Psi_{max}]$, where $\Psi_{min}$ and $\Psi_{max}$ are the minimum and maximum values of $\Psi$, respectively. To connect the velocity and the space domain, let there be a deterministic relation

$$\hat{u} = t(\Psi), t \text{ being any function.} \qquad (10)$$

$\Psi(Y, Z)$ represents an arbitrary point in a channel cross section. Assume that the values of $\Psi$ in between $(\Psi_{min}, \Psi_{max})$ are equally likely and are hence assumed to follow a uniform probability distribution as follows

$$\check{f}(\Psi) = \frac{1}{\Psi_{max} - \Psi_{min}} \qquad (11)$$

Then, using the relation given by Eq. (10), distribution of $\hat{u}$ can be written as

$$f(\hat{u}) = \check{f}(\Psi) \left|\frac{d\Psi}{d\hat{u}}\right| = \frac{1}{\Psi_{max} - \Psi_{min}} \frac{d\Psi}{d\hat{u}} \qquad (12)$$

Furthermore, the CDF can be recovered as follows

$$F(\hat{u}) = Prob(\hat{U} \leq \hat{u}) = \int_0^{\hat{u}} f(\hat{u})d\hat{u} = \int_{t^{-1}(0)}^{t^{-1}(\hat{u})} f(t(\Psi))\frac{d\hat{u}}{d\Psi}d\Psi = \int_{\Psi_{min}}^{\Psi} \check{f}(\Psi)d\Psi$$

$$= \frac{\Psi - \Psi_{min}}{\Psi_{max} - \Psi_{min}} \qquad (13)$$

It may be noted from Eq. (13) that if the velocity is randomly sampled over the entire channel cross-section then the area enclosed by the channel boundary and an isovel (velocity contour) on which the velocity is $\hat{u}$ divided by the total area of the cross section is equivalent to the probability of velocity being less than or equal to $\hat{u}$. For wide open channels where velocity varies only in the vertical direction, $\Psi_{min} = 0$, $\Psi_{max} = 1$ and $\Psi$ can be well approximated as $\Psi = y/D$, $\Psi_{min} = 0$, and $\Psi_{max} = 1$ (*Chiu 1989*). Accordingly, Eq. (13) takes on the form

$$F(\hat{u}) = \frac{y}{D} \qquad (14)$$

### 2.3. Derivation of Velocity Equation

Equating the entropy-based CDF Eq. (8) and the hypothesized CDF in the space domain Eq. (13) and doing some algebraic arrangement, the velocity equation can be obtained as

$$\hat{u} = -\frac{\lambda_0 + \frac{1}{q-1}}{\lambda_1} + \frac{1}{\lambda_1}\left[\left(\lambda_0 + \frac{1}{q-1}\right)^{\frac{q}{q-1}} + \lambda_1 \left(\frac{q}{q-1}\right)^{\frac{q}{q-1}} \frac{\Psi - \Psi_{min}}{\Psi_{max} - \Psi_{min}}\right]^{\frac{q-1}{q}} \qquad (15)$$

It may be noted that the derived velocity profile depends on three parameters, namely, the Lagrange multipliers ($\lambda_0$ and $\lambda_1$) and the entropy index.

### 2.4. Combining the Lagrange Multipliers

To simplify the derived equations (*Chiu 1987*, *Luo and Singh 2011*, *Cui and Singh 2014*), let us introduce the following parameter called entropy parameter combining the Lagrange multipliers as follows

$$G = \frac{\lambda_1}{\lambda_0 + \lambda_1 + \frac{1}{q-1}} \qquad (16)$$

From Eq. (16), one obtains

$$\lambda_0 + \frac{1}{q-1} = \lambda_1 \left(\frac{1-G}{G}\right) \qquad (17)$$

Also, the first constraint Eq. (6) can be rearranged as

$$\lambda_1^{\frac{1}{q-1}} = \frac{\left(\frac{q}{q-1}\right)^{\frac{q}{q-1}}}{\left(\frac{1}{G}\right)^{\frac{q}{q-1}} - \left(\frac{1-G}{G}\right)^{\frac{q}{q-1}}} \qquad (18)$$

Using Eqs. (16)-(18), the velocity distribution can be written in terms of two parameters as follows

$$\hat{u} = \frac{G-1}{G} + \left[\left(\frac{1-G}{G}\right)^{\frac{q}{q-1}} + \left\{\left(\frac{1}{G}\right)^{\frac{q}{q-1}} - \left(\frac{1-G}{G}\right)^{\frac{q}{q-1}}\right\} \frac{\Psi - \Psi_{min}}{\Psi_{max} - \Psi_{min}}\right]^{\frac{q-1}{q}} \qquad (19)$$

For convenience, let us introduce $\alpha_1 = \frac{q}{q-1}$ and $\alpha_2 = \frac{1-G}{G}$ for which Eq. (19) is further simplified to

$$\hat{u} = -\alpha_2 + \left[\alpha_2^{\alpha_1} + \{(1+\alpha_2)^{\alpha_1} - \alpha_2^{\alpha_1}\} \frac{\Psi - \Psi_{min}}{\Psi_{max} - \Psi_{min}}\right]^{\frac{1}{\alpha_1}} \qquad (20)$$

### *2.5. Estimation of the Parameters*

It can be seen that Eq. (20) contains parameters $\alpha_1$ and $\alpha_2$ the determination of which is essentially needed for the assessment of the velocity profile. One parameter is based on the Tsallis entropy index and the other is the defined entropy parameter. *Luo and Singh (2011)* and *Singh and Luo (2011)* considered some test values for the entropy index $q$ and based on the data-fitting procedure they found the best value to be 3/4 and 2 for 1D and 2D velocity distributions, respectively. Extending their work by introducing some empirical parameters in the hypothesized CDF, *Cui and Singh (2013, 2014)* determined the best choice for $q$ to be 3 for both 1D and 2D velocity distributions. However, the approaches proposed by them are only a heuristic way and based on data, and also provide no physical justification for the entropy index. Here, we provide physical justification to the entropy index in modelling velocity in open channels based on the method of moments. Using Eqs. (16)-(18) and introducing parameters $\alpha_1$ and $\alpha_2$, the PDF given by Eq. (5) can be written as

$$f(\hat{u}) = \frac{\alpha_1}{(1+\alpha_2)^{\alpha_1} - (\alpha_2)^{\alpha_1}} (\alpha_2 + \hat{u})^{\alpha_1 - 1} \qquad (21)$$

Eq. (21) is a power-law distribution obtained from the maximization of Tsallis entropy subject to the total probability rule and mass conservation constraint. For estimating parameters $\alpha_1$ and $\alpha_2$ that characterize the PDF Eq. (21), one may use the method of maximum likelihood estimation (MLE). However, it should be noted that the parameters are dependent on each other as can be noticed from Eq. (16). To that end, the MLE technique would require an explicit expression for their derivative(s). Because of these limitations, the application of this technique may be difficult. Therefore, we proceed with the method of moments which is simpler to determine the estimators.

Here we need to determine the first two moments for estimating the two unknown parameters $\alpha_1$ and $\alpha_2$ that characterize the PDF $f(\hat{u}; \alpha_1, \alpha_2)$ of velocity. After performing the integrals, the first two moments can be expressed as follows

$$\mu_1 = E[\widehat{U}] = \bar{\hat{u}} = \int_0^1 \hat{u} f(\hat{u}) d\hat{u} = \frac{\alpha_1}{\alpha_1 + 1} \frac{(1+\alpha_2)^{\alpha_1+1} - (\alpha_2)^{\alpha_1+1}}{(1+\alpha_2)^{\alpha_1} - (\alpha_2)^{\alpha_1}} - \alpha_2 \qquad (22)$$

$$\mu_2 = E[\widehat{U}^2] = \overline{\hat{u}^2} = \beta \bar{\hat{u}}^2 = \int_0^1 \hat{u}^2 f(\hat{u}) d\hat{u}$$

$$= \frac{\alpha_1}{\alpha_1 + 2} \frac{(1+\alpha_2)^{\alpha_1+2} - (\alpha_2)^{\alpha_1+2}}{(1+\alpha_2)^{\alpha_1} - (\alpha_2)^{\alpha_1}} - \frac{2\alpha_1 \alpha_2}{\alpha_1 + 1} \frac{(1+\alpha_2)^{\alpha_1+1} - (\alpha_2)^{\alpha_1+1}}{(1+\alpha_2)^{\alpha_1} - (\alpha_2)^{\alpha_1}}$$
$$+ \alpha_2^{\,2} \qquad (23)$$

where E is the expectation operator and $\beta$ is the momentum diffusion coefficient or *Boussinesq coefficient* equal to $\overline{\hat{u}^2}/\bar{\hat{u}}^2$ (*Chow 1959*). It is pertinent to mention that Eq. (22) is the mass conservation constraint that we have taken into account while maximizing the entropy function. On the other hand, the second order moment considered in Eq. (23) is based on the conservation of momentum as conveyed in the following (*Chiu 1989*). The momentum transport (per unit time) by the flow through a cross section is $\rho \int_A u^2 dA$, so

$$\rho \int_A u^2 dA = \rho u_{max}^2 \int_A \hat{u}^2 dA = \rho A u_{max}^2 \overline{\hat{u}^2}(= \beta \rho A u_{max}^2 \bar{\hat{u}}^2) = \rho A u_{max}^2 \int_0^1 \hat{u}^2 f(\hat{u})d\hat{u} \qquad (24)$$

Now consider a random sample of velocity data of size $n$, namely $\hat{u}_1, \hat{u}_2, \ldots, \hat{u}_n$. Then the sample moments can be estimated as $\widehat{\mu_1} = \frac{1}{n}\sum_{i=1}^n \hat{u}$ and $\widehat{\mu_2} = \beta\left(\frac{1}{n}\sum_{i=1}^n \hat{u}\right)^2$. Hence, the estimators for $\alpha_1$ and $\alpha_2$ denoted by $\widehat{\alpha_1}$ and $\widehat{\alpha_2}$, based on the method of moments, constitute the solution of the given system of non-linear equations

$$\widehat{\mu_1} = g_1(\widehat{\alpha_1}, \widehat{\alpha_2}) \qquad (25)$$

$$\widehat{\mu_2} = g_2(\widehat{\alpha_1}, \widehat{\alpha_2}) \qquad (26)$$

where $g_i(\blacksquare)$ for $i = 1,2$ are the right-hand sides of Eqs. (22) and (23), respectively. It is concluded that while modelling velocity in open channels using Tsallis entropy, the present work provides a physical justification for the entropy index $q$. To be more specific, the entropy index along with the defined entropy parameter depends on the normalized mean velocity as well as on momentum coefficient. To solve Eqs. (25) and (26), an expression for $\beta$ is needed. For that purpose, the following expressions are considered.

$$\beta = 1 + \left(\frac{u_{max}}{\bar{u}} - 1\right)^2 \qquad (27)$$

and

$$\beta = \frac{(\exp(M) - 1)[(M^2 - 2M + 2)\exp(M) - 2]}{[(M-1)\exp(M) + 1]^2} \qquad (28)$$

where the entropy parameter $M$ is given by $\frac{\bar{u}}{u_{max}} = \frac{\exp(M)}{\exp(M)-1} - \frac{1}{M}$. Eq. (27) was given by *Chow (1959)*, developed on the basis of the logarithmic velocity distribution. On the other hand, *Chiu and Hsu (2006)* derived Eq. (28) based on the entropy concept. With the values of $\bar{\hat{u}}$ and $\beta$, Eqs. (25) and (26) can be solved together numerically to obtain the estimates $\widehat{\alpha_1}$ and $\widehat{\alpha_2}$. Newton's method may lack in efficiency due to the choice of the initial guess. Therefore, we used the *trust-region* algorithm to solve the equations, which is a simple yet powerful optimization technique (*Conn et al. 2000*). MatLab function '*fsolve*' contains three algorithms for solving a system of equations, namely, trust-region, trust-region dogleg, and Levenberg-Marquardt. In the present study, the trust-region dogleg scheme was used as it is the most efficient as compared to the others.

## 3. Results and Discussion

The present work provides a physical interpretation of the Tsallis entropy index for modelling velocity in open channel flow. In addition, to assess how the proposed approach improves the velocity equation, the derived equation is compared with the existing one considering different data sets for both one- and two-dimensional distributions of streamwise velocity. Furthermore, a quantitative evaluation is done for the velocity models using the relative percentage error (RE) and root-mean-square error (RMSE).

### *3.1. Experimental and Field Data Considered*

For the validation of velocity equation in a wide channel where velocity varies in the vertical direction having the maximum velocity at the water surface, mostly cited laboratory data of *Einstein and Chien (1955)* and field data of *Davoren (1985)* were considered. *Einstein and Chien (1955)* measured data for both clear water and sediment-laden flow up to 50% of the flow depth starting from the channel bed. A detailed description can be found in *Einstein and Chien (1955)*. Field data of *Davoren (1985)* was for a river with a live bed. The flow velocities were measured downstream from a hydropower plant, which resulted in a steady uniform flow over an appreciable period of time. On the other hand, for velocity with dip-phenomenon and two-dimensional distribution of velocity, laboratory data of *Wang and Qian (1989)* and field data of *Moramarco and Singh (2001)* were considered, respectively. For details one can refer to the cited literature.

### *3.2. Measurement of Error*

To get a quantitative idea about the validity of velocity equations, RE and RMSE were calculated as follows

$$RE = \frac{1}{N}\sum_{i=1}^{N}\left|\frac{V_c(i) - V_o(i)}{V_o(i)}\right| \qquad (29)$$

$$RMSE = \sqrt{\frac{1}{N}\sum_{i=1}^{N}\left(\frac{V_c(i) - V_o(i)}{V_o(i)}\right)^2} \qquad (30)$$

where $N$ is the total number of node points, $V_c(i)$ and $V_0(i)$ are the computed and observed values at $i$-th data point, respectively.

### 3.3. Assessment of Velocity Profile

In this section, the derived velocity equation was compared with selected data sets and also with the models of *Singh and Luo (2011)* and *Luo and Singh (2011)*. Their models can be achieved by solving the system of Eqs. (6) and (7) after putting $q = 3/4$ and 2 for velocity in wide and narrow channels, respectively.

### 3.3.1. Velocity in Wide Channels

To compare the model in the case of flow in wide channels, Run C3 (clear water flow) and Run S5 (sediment-laden flow) of *Einstein and Chien (1955)* data, and Run 1 and 10 of field data from *Davoren (1985)* were considered. The characteristics of data are provided in Table 1. The proposed model was assessed for two cases: considering $\beta$ from Eq. (27) and from Eq. (28). Fig. 1 compares the proposed model and the model of *Singh and Luo (2011)* with Run 1 and 10 of *Davoren (1985)* data. It can be noticed from the figure that the proposed model for both cases significantly improves the prediction of velocity throughout the water column as compared to the *Singh and Luo (2011)* model. In Fig. 2, the models were compared for data from clear water as well as sediment-laden flow of *Einstein and Chien (1955)*. The superiority of the present model was observed here also, especially near the channel bed. The absolute relative error (RE) and the root-mean-square error (RMSE) were calculated for the models and are presented in Table 2. It can be noticed from the table that the present model mimiced the data in a much better way than did the model of *Singh and Luo (2011)*. Furthermore, it was observed that the prediction accuracies of the present models for the two cases of calculation of $\beta$ were almost the same.

Table 1: Characteristics of selected velocity data (for wide channels).

| Data Set | | $\bar{u}$ (m/s) | $u_{max}$ (m/s) | $M$ | Momentum coefficient $\beta$ | | $\beta$ from Eq. (27) | | $\beta$ from Eq. (28) | | *Singh and Luo (2011)* ($q = 3/4$) | |
|---|---|---|---|---|---|---|---|---|---|---|---|---|
| | | | | | Eq. (27) | Eq. (28) | $\alpha_1$ | $\alpha_2$ | $\alpha_1$ | $\alpha_2$ | $\lambda_0$ | $\lambda_1$ |
| | Run 1 | 2.285 | 2.807 | 5.225 | 1.052 | 1.047 | 156.933 | 28.851 | 156.894 | 29.140 | -1.821 | 3.944 |

| | | | | | | | | | | | |
|---|---|---|---|---|---|---|---|---|---|---|---|
| *Davoren (1985)* | Run 10 | 2.258 | 2.790 | 5.080 | 1.055 | 1.050 | 156.826 | 29.646 | 156.777 | 29.981 | -1.729 | 3.836 |
| *Einstein and Chien (1955)* | Run C3 | 1.960 | 2.288 | 6.928 | 1.028 | 1.027 | 145.431 | 20.054 | 155.411 | 21.548 | -2.885 | 5.158 |
| | Run S5 | 2.226 | 2.802 | 4.645 | 1.067 | 1.058 | 156.427 | 32.242 | 156.336 | 32.767 | -1.457 | 3.514 |

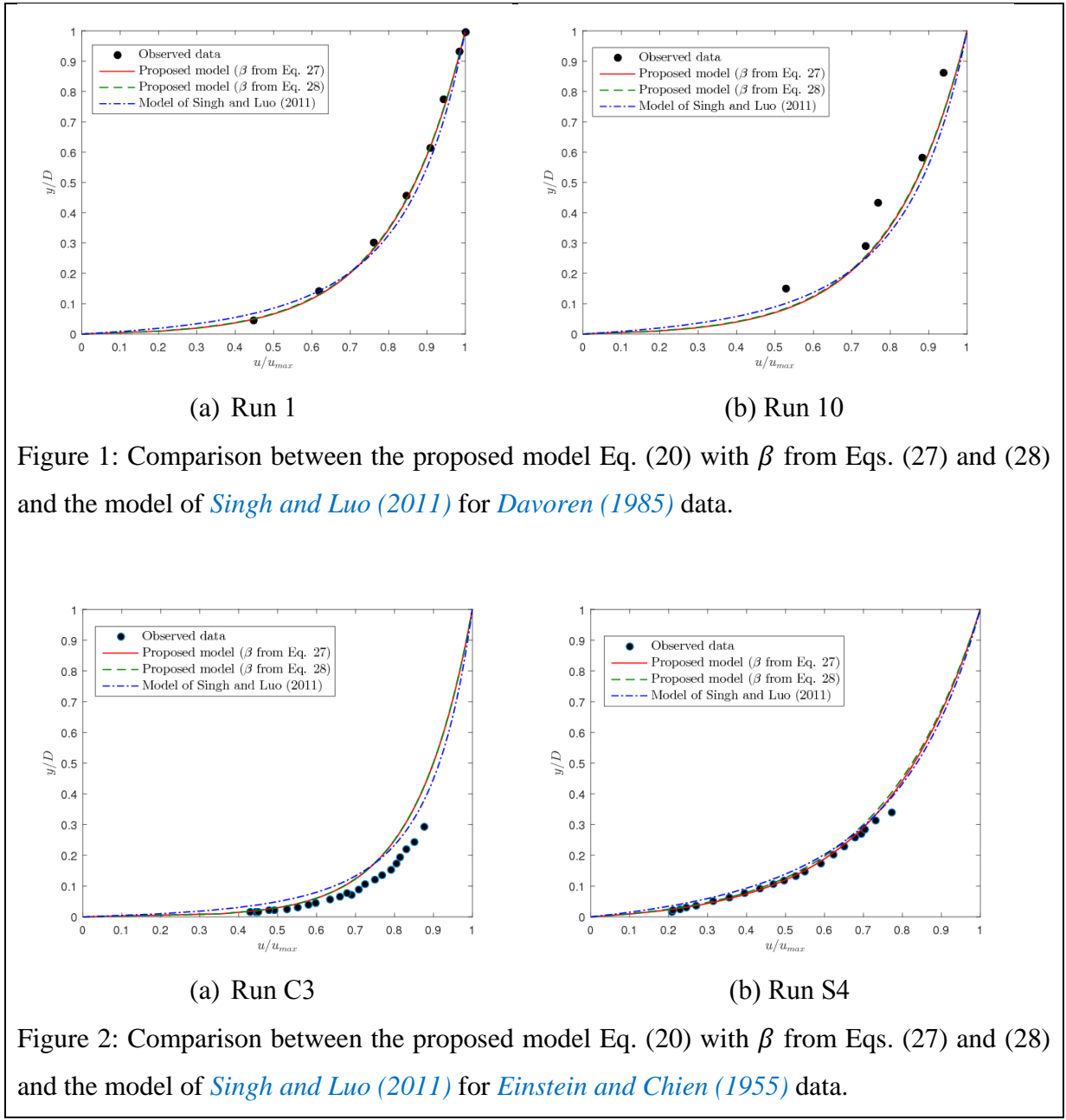

(a) Run 1  (b) Run 10

Figure 1: Comparison between the proposed model Eq. (20) with $\beta$ from Eqs. (27) and (28) and the model of *Singh and Luo (2011)* for *Davoren (1985)* data.

(a) Run C3  (b) Run S4

Figure 2: Comparison between the proposed model Eq. (20) with $\beta$ from Eqs. (27) and (28) and the model of *Singh and Luo (2011)* for *Einstein and Chien (1955)* data.

Table 2: RE and RMSE values for the models.

| Data Set | | Proposed Model ($\beta$ from Eq. 27) | | Proposed Model ($\beta$ from Eq. 28) | | Model of *Singh and Luo (2011)* | |
|---|---|---|---|---|---|---|---|
| | | RE | RMSE | RE | RMSE | RE | RMSE |
| *Davoren (1985)* | Run 1 | 0.0122 | 0.0173 | 0.0123 | 0.0192 | 0.0377 | 0.0755 |
| | Run 10 | 0.0634 | 0.0941 | 0.0611 | 0.0910 | 0.0658 | 0.0879 |
| *Einstein and Chien (1955)* | Run C3 | 0.0696 | 0.0705 | 0.0714 | 0.0722 | 0.1815 | 0.2165 |
| | Run S5 | 0.0359 | 0.0655 | 0.0484 | 0.0828 | 0.1243 | 0.1765 |

### 3.3.2. Velocity in Narrow Channels

Run 1 (clear water flow) and 31 (sediment-laden flow) of *Wang and Qian (1989)* data where maximum velocity occurred below the water surface and three verticals $z = -16.64$ m, $z = 0$ m and $z = 16.64$ m located at P. Nuovo gauged section (*Moramarco and Singh 2001*) for two-dimensional distribution were considered for the comparison. The characteristics of data sets are provided in Table 3. Fig. 3 compares the proposed model and the model of *Luo and Singh (2011)* for Run 1 and 31 of *Davoren (1985)* data. The RE and RMSE values were calculated and are presented in Table 4. It can be observed from the table that the present model predicted the data better in the case of Run 1 while for other data set the model of *Luo and Singh (2010)* performed better. However, the proposed model measured the data in the near-bed region more efficiently as can be seen from the figure. In Fig. 4(a), the vertical velocity profiles at $z = -16.64$ m and $z = 0$ m of P. Nuovo gauged section (*Moramarco and Singh 2001*) were plotted and compared with the velocity models. It can be seen from the figure and Table 4 that the present model significantly improved the velocity profile as compared to the model of *Luo and Singh (2011)*. Moreover, the model where $\beta$ was computed from Eq. (27) was superior to the other and with this model, the velocity contours are plotted in Fig. 4(b). It is observed from the figure that the derived model can measure the velocity well along the cross-section of the channel.

Table 3: Characteristics of selected velocity data (for narrow channels).

| Data Set | | $\bar{u}$ (m/s) | $u_{max}$ (m/s) | $h$ (m) | $M$ | Momentum coefficient $\beta$ | | $\beta$ from Eq. 27 | | $\beta$ from Eq. 28 | | *Luo and Singh (2011)* ($q = 2$) | |
|---|---|---|---|---|---|---|---|---|---|---|---|---|---|
| | | | | | | Eq. (27) | Eq. (28) | $\alpha_1$ | $\alpha_2$ | $\alpha_1$ | $\alpha_2$ | $\lambda_0$ | $\lambda_1$ |
| | Run 1 | 1.720 | 2.116 | -0.04 | 5.184 | 1.053 | 1.048 | 156.904 | 29.071 | 156.862 | 29.377 | -2.753 | 7.506 |

| Wang and Qian (1989) | Run 31 | 1.641 | 2.132 | -0.03 | 4.023 | 1.090 | 1.073 | 158.917 | 37.535 | 158.698 | 38.549 | -2.236 | 6.472 |
| Moramarco and Singh (2001) | Three verticals at Pt. Nuovo | 1.684 | 2.597 | -2.60 | 1.886 | 1.294 | 1.167 | 140.149 | 60.565 | 120.047 | 62.631 | -0.783 | 3.565 |

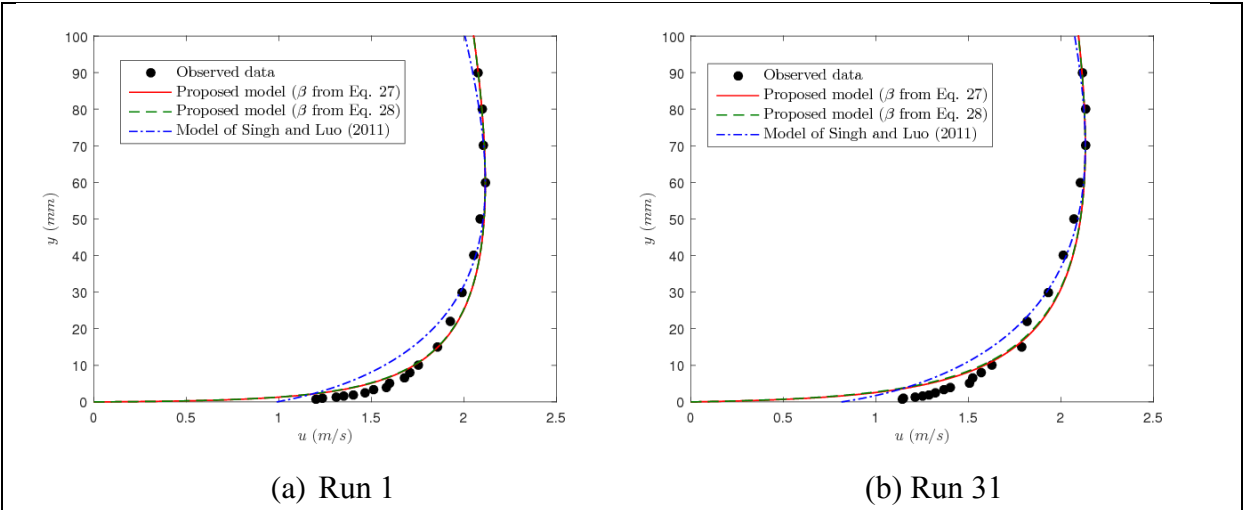

(a) Run 1          (b) Run 31

Figure 3: Comparison between the proposed model Eq. (20) with $\beta$ from Eqs. (27) and (28) and the model of *Luo and Singh (2011)* for *Wang and Qian (1989)* data.

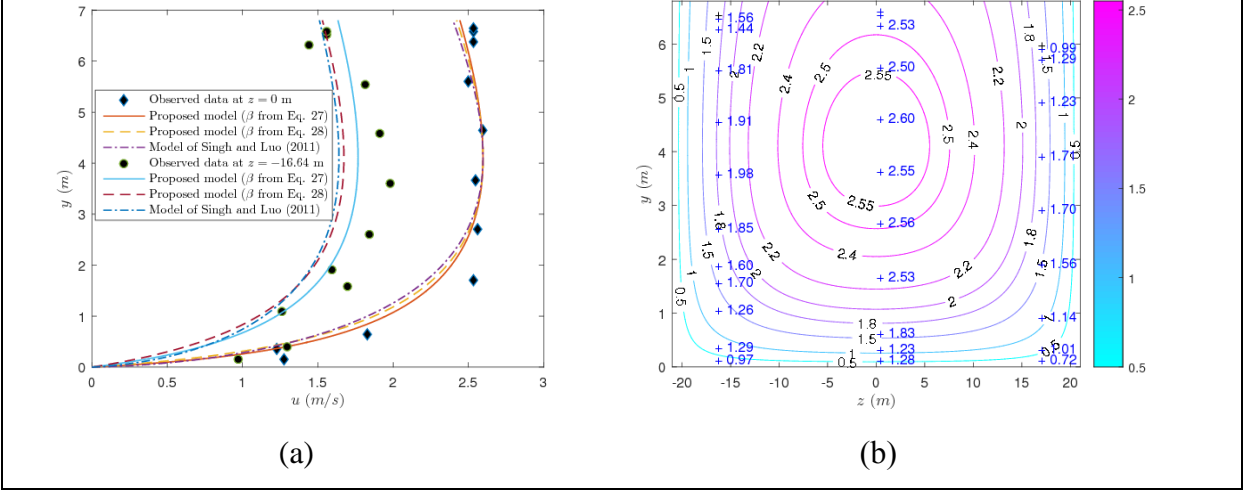

(a)          (b)

Figure 4: Comparison between the proposed model Eq. (20) with $\beta$ from Eqs. (27) and (28) and the model of *Luo and Singh (2011)* for *Moramarco and Singh (2001)* data: (a) along two verticals, and (b) contour lines. [texts with '+' inside the figures indicate the observed data]

Table 4: RE and RMSE values for the models.

| Data Set | | Proposed Model ($\beta$ from Eq. 27) | | Proposed Model ($\beta$ from Eq. 28) | | Model of *Luo and Singh* (2011) | |
|---|---|---|---|---|---|---|---|
| | | RE | RMSE | RE | RMSE | RE | RMSE |
| *Wang and Qian (1989)* | Run 1 | 0.0867 | 0.1311 | 0.0896 | 0.1354 | 0.0907 | 0.1141 |
| | Run 31 | 0.1474 | 0.2214 | 0.1566 | 0.2332 | 0.1076 | 0.1374 |
| *Moramarco and Singh (2001)* | $z = -16.64$ m | 0.1541 | 0.2400 | 0.1862 | 0.2763 | 0.1721 | 0.2355 |
| | $z = 0$ m | 0.0731 | 0.1448 | 0.0979 | 0.1771 | 0.0915 | 0.1529 |
| | $z = 16.64$ m | 0.3133 | 0.4109 | 0.3053 | 0.3881 | 0.2652 | 0.3363 |

## 4. Conclusions

The following conclusions can be drawn from the present work:

1. The present study revisits the velocity distribution in open channels using Tsallis entropy with special attention to the role of Tsallis entropy index. Using the method of moments, it is found that the entropy index together with the defined parameter depends on the normalized mean velocity and momentum coefficient.

2. The entropy parameter and the entropy index are determined simultaneously by solving a system of non-linear equations that arise through the application of the method of moments. The second order statistical moment based on the conservation of momentum is considered.

3. Relevant sets of laboratory and field data are considered for assessing the velocity profile. In addition to the physical meaning of the entropy index, it is observed that the derived velocity profile significantly improves for both wide and narrow channels.

4. A quantitative assessment is also done for the models through the computations of absolute relative error and root-mean-square error. The proposed approach may be used for other open channel flow problems, such as sediment concentration and shear stress distribution, hindered settling velocity, bed-load layer thickness, etc.

**ACKNOWLEDGMENTS**

The first two authors are thankful to the Science and Engineering Research Board (SERB), Department of Science and Technology (DST), Govt. of India for providing the financial support through the research project with no.: **SERB/F/4873/2018-2019, Dated 24 July, 2018**.

**References:**

1. Chiu, C. L. (1987). Entropy and probability concepts in hydraulics. *Journal of Hydraulic Engineering*, *113*(5), 583-599.

2. Chiu, C. L. (1989). Velocity distribution in open channel flow. *Journal of Hydraulic Engineering*, *115*(5), 576-594.

3. Chiu, C. L., and Hsu, S. M. (2006). Probabilistic approach to modeling of velocity distributions in fluid flows. *Journal of Hydrology*, *316*(1-4), 28-42.

4. Chow, V.T. (1959). *Open-channel hydraulics* (Vol. 1). New York: McGraw-Hill.

5. Conn, A. R., Gould, N. I., and Toint, P. L. (2000). *Trust region methods* (Vol. 1). Siam.

6. Conroy, J. M., and Miller, H. G. (2015). Determining the Tsallis parameter via maximum entropy. *Physical Review E*, *91*(5), 052112.

7. Cui, H., and Singh, V. P. (2013). Two-dimensional velocity distribution in open channels using the Tsallis entropy. *Journal of Hydrologic Engineering*, *18*(3), 331-339.

8. Cui, H., and Singh, V. P. (2014). One-dimensional velocity distribution in open channels using Tsallis entropy. *Journal of Hydrologic Engineering*, *19*(2), 290-298.

9. Davoren, A. (1985). *Local scour around a cylindrical bridge pier*. Hydrology Centre, Ministry of Works and Development for the National Water and Soil Conservation Authority.

10. Einstein, H. A., and Chien, N. (1955). Effects of heavy sediment concentration near the bed on velocity and sediment distribution, MRD Sediment Ser. 8. *Univ. of Calif., Berkeley*.

11. Keylock, C. J. (2005). Describing the recurrence interval of extreme floods using nonextensive thermodynamics and Tsallis statistics. *Advances in water resources*, *28*(8), 773-778.

12. Koutsoyiannis, D. (2005a). Uncertainty, entropy, scaling and hydrological stochastics. 1. Marginal distributional properties of hydrological processes and state scaling. *Hydrological Sciences Journal*, *50*(3), 381-404.

13. Koutsoyiannis, D. (2005b). Uncertainty, entropy, scaling and hydrological stochastics. 2. Time dependence of hydrological processes and time scaling. *Hydrological Sciences Journal*, *50*(3), 405-426.

14. Kumbhakar, M., Kundu, S., and Ghoshal, K. (2017). Hindered settling velocity in particle-fluid mixture: a theoretical study using the entropy concept. *Journal of Hydraulic Engineering*, *143*(11), 06017019.

15. Kumbhakar, M., Kundu, S., and Ghoshal, K. (2018). An explicit analytical expression for bed-load layer thickness based on maximum entropy principle. *Physics Letters A*, *382*(34), 2297-2304.

16. Kundu, S. (2017). Prediction of velocity-dip-position over entire cross section of open channel flows using entropy theory. *Environmental Earth Sciences*, *76*(10), 363.

17. Luo, H., and Singh, V. P. (2011). Entropy theory for two-dimensional velocity distribution. *Journal of Hydrologic Engineering*, *16*(4), 303-315.

18. Lyra, M. L., and Tsallis, C. (1998). Nonextensivity and multifractality in low-dimensional dissipative systems. *Physical review letters*, *80*(1), 53.

19. Moramarco, T., and Singh, V. P. (2001). Simple method for relating local stage and remote discharge. *Journal of Hydrologic Engineering*, *6*(1), 78-81.

20. Plastino, A. R. P. A., and Plastino, A. R. (1999). Tsallis Entropy and Jaynes' information theory formalism. *Brazilian Journal of Physics*, *29*(1), 50-60.

21. Singh, V. P., and Luo, H. (2011). Entropy theory for distribution of one-dimensional velocity in open channels. *Journal of Hydrologic Engineering*, *16*(9), 725-735.

22. Singh, V. P. (2014). *Entropy theory in hydraulic engineering: an introduction*. American Society of Civil Engineers (ASCE).

23. Singh, V. P. (2016). *Introduction to Tsallis entropy theory in water engineering*. CRC Press.

24. Tsallis, C. (1988). Possible generalization of Boltzmann-Gibbs statistics. *Journal of statistical physics*, *52*(1-2), 479-487.

25. Tsallis, C. (1999). Nonextensive statistics: theoretical, experimental and computational evidences and connections. *Brazilian Journal of Physics*, *29*(1), 1-35.

26. Wang, X., and Qian, N. (1989). Turbulence characteristics of sediment-laden flow. *Journal of hydraulic Engineering*, *115*(6), 781-800.